# Three dimensional optical manipulation and structural imaging of soft materials by use of laser tweezers and multimodal nonlinear microscopy

**Rahul P. Trivedi[1], Taewoo Lee[1], Kris A. Bertness[2], and Ivan I. Smalyukh[1,3] \***

[1]*Department of Physics and Liquid Crystal Materials Research Center, University of Colorado, Boulder, CO 80309, USA*
[2]*National Institute of Standards and Technology, Boulder, Colorado 80305, USA*
[3]*Renewable & Sustainable Energy Institute, University of Colorado and National Renewable Energy Laboratory, Boulder, CO 80309, USA*
*\*Email: Ivan.Smalyukh@Colorado.Edu*

**Abstract:**

We develop an integrated system of holographic optical trapping and multimodal nonlinear microscopy and perform simultaneous three-dimensional optical manipulation and non-invasive structural imaging of composite soft-matter systems. We combine different nonlinear microscopy techniques such as coherent anti-Stokes Raman scattering, multi-photon excitation fluorescence and multi-harmonic generation, and use them for visualization of long-range molecular order in soft materials by means of their polarized excitation and detection. The combined system enables us to accomplish both, manipulation in composite soft materials such as colloidal inclusions in liquid crystals as well as imaging of each separate constituents of the composite material in different nonlinear optical modalities. We also demonstrate optical generation and control of topological defects and simultaneous reconstruction of their three-dimensional long-range molecular orientational patterns from the nonlinear optical images.

## 1. Introduction

Soft materials often combine properties such as fluidity, intrinsic to conventional liquids, with the properties and structural features typically associated with solid crystals, such as elasticity, orientational and positional order, anisotropy, and topological defects. Characterization of physical properties and experimental study of various physical phenomena in such materials is a challenging task, as all properties depend on the details of structural organization, presence of defects, etc. Liquid crystals (LCs) are a classic example of soft materials, typically possessing long-range orientational order combined with varying degree of positional order [1]. The simplest type of LCs, the so-called nematic LC, is composed of rod-shaped molecules whose local average orientation is described by the molecular director, $\mathbf{n(r)}$, which is also the local optical axis of these materials. LCs have found both fundamental and technological importance, e.g., in display and other electro-optic devices, and are also significant from biological standpoint as lipids, viruses, and membranes form LC phases [2-6]. Introducing chirality in nematic LCs enriches the phase behavior by resulting in chiral nematic phases, viz., cholesteric and blue phases [1]. In the cholesteric LC, the molecules show a uniform twist along a helical axis, giving rise to a periodic lamellar-like ground state.

The inherent anisotropy of LC materials imparts unique properties to LC-based colloidal dispersions [7,8]. The elasticity- and defect-mediated interactions between colloids in LCs are markedly different from those between colloids dispersed in isotropic media, allowing for novel ways of nano-scale and micro-scale colloidal self-assembly. Optical manipulation, which has already marked its importance in the field of colloidal science [9] and biology [10], has attained a place of pivotal interest in probing these so-called "LC colloids" [11-17], and multitudes of other LC systems such as defect-structures [17-21], Langmuir monolayers [22] and LC droplets [23, 24]. Optical trapping has proved to be a valuable tool in determining the nature of interaction forces [11-13], characterizing viscous drag on colloidal inclusions in LCs [25, 26], controlling dynamics of trapped particles by us of direction-dependent optical trapping forces in LCs [15], probing defect topology [16,21], and characterizing line tension of disclination defects [18, 19, 21], etc. Optical manipulation in LCs is not restricted to tweezing of foreign inclusion but extends to manipulation of the director field **n(r)** with a focused laser beam – leading to generation of localized defect structures [27] which, in turn, are amenable to optical tweezing.

While performing optical manipulation in LCs, it is imperative that the changes in **n(r)** effected by inclusion of colloids, or by application of external optical/electric fields, be imaged to elucidate the underlying physical phenomena. For example, characterization of anisotropic viscoelastic properties of LCs using active microrheology assisted by laser tweezers is complicated by the fact that the probe particles typically induce distortions of long-range molecular alignment and topological defects [15, 25, 26], which need to be visualized to provide a deeper understanding of the nature of ensuing colloidal interactions. The conventional technique of optical polarizing microscopy reveals the depth-averaged **n(r)** in the lateral plane perpendicular to the microscope's optical axis; and it is often impossible to deduce the full three-dimensional (3D) **n(r)**. Combination of optical manipulation with 3D imaging technique such as fluorescence confocal polarizing microscopy (FCPM) [28-31] has enabled substantial progress in LC imaging. This approach however requires the LC to be doped with fluorescent dyes whose molecules align along the local **n(r)** without distorting it. For certain imaging applications such as imaging of some biological systems or characterization of pre-packaged devices, introduction of dyes is precluded, as it may either be obtrusive to their natural functioning or undesired for their successive application. Hence, several nonlinear optical microscopy techniques have been used for imaging of **n(r)** in LCs: viz., coherent anti-Stokes Raman scattering (CARS) [32-34], four-wave mixing [35], second-harmonic generation (SHG) [36], third-harmonic generation [37], three-photon excitation fluorescence (3PEF) [38], etc. These techniques, and also multimodal nonlinear optical polarizing microscopy (MNOPM) that combines multiple nonlinear optical imaging modalities [39], address this issue by directly probing the ordered arrangement of constituent molecules or building-blocks of the LCs instead of relying on additives such as dyes. However, simultaneous implementation of these 3D imaging techniques with 3D optical manipulation has never been demonstrated. We show here the simultaneous use of holographic optical trapping (HOT) with the MNOPM imaging system. The MNOPM includes imaging modalities such as CARS, multi-photon excitation fluorescence (in particular, we show imaging performed in 3PEF), and multi-harmonic generation (with the example of imaging in SHG). This integrated system gives capabilities of non-contact 3D optical manipulation along with non-invasive, label-free, high resolution and orientation sensitive 3D imaging.

## 2. Experimental

### 2.1 Optical setup:

The schematic diagram of the integrated MNOPM and HOT system is shown in Fig. 1. The two systems are coupled to an inverted microscope (IX-81, Olympus [40]). A tunable (680-1080 nm) Ti-Sapphire oscillator (140 fs, 80 MHz, Chameleon Ultra II, Coherent) is used as the primary excitation source for the nonlinear optical imaging modalities. For CARS imaging, the femtosecond pulse at 780 nm is split into two arms: the pump/probe beam and the other beam which synchronously pumps a highly nonlinear, polarization maintaining photonic crystal fiber (PCF, FemtoWHITE-800, NKT photonics). The broadband Stokes pulse is derived from the supercontinuum output of the PCF (spanning a range from ~550nm to ~1600 nm). The two beams are recombined at a filter (BLP01-785R-25, Semrock Inc.) that spatially overlaps the pump/probe pulse (at 780 nm) and the broadband Stokes pulse

(consisting of the part of the supercontinuum above 785 nm). The temporal overlap of the two pulses is ensured by adjusting delay lines in each arm, before directing them into the galvanomirror-based scanning unit (Fluoview FV300, Olympus). For SHG and 3PEF, the pulse is shunted directly into the scanning unit. The excitation light is focused into the sample by use of a high numerical aperture (NA) objective lens (100x, oil-immersion, NA=1.4, Olympus) and collected by another objective (60x, oil-immersion, NA=1.42, Olympus) in forward mode or the same one for detection in epi-mode. The nonlinear signals are detected with photomultiplier tubes (H5784-20, Hamamatsu) after passing through a series of dichroic mirrors and bandpass filters. A Faraday isolator is used to protect the Ti:Sapphire laser from the back-reflection of the PCF, and a combination of a half-wave plate and a Glan laser polarizer allows us to control power and polarization of the laser. The pulse is passed through a dual prism dispersion compensator before it is split. A laser line filter (LL01-780, Semrock Inc) is used to reduce the spectral bandwidth of pump/probe pulses for suppressing the non-resonant background signal in the CARS imaging mode. All spectra are measured at forward detection position by an optical-fiber – coupled spectrometer (USB2000-FLG, Ocean Optics).

The HOT system (see Fig. 1) is built around a reflective, electrically addressed, phase-only, liquid crystal spatial light modulator (SLM, Boulder Nonlinear Systems) with 512×512 pixels, each of size 15×15 μm. Refresh rate of 15 Hz for the computer-generated holograms supplied to the SLM ensures real-time manipulation. The trapping system employs a continuous-wave Ytterbium-doped fiber laser at 1064 nm (YLM-10, IPG Photonics). A Glan laser polarizer and a half-wave plate control the power and the polarization of the beam incident on the SLM (the polarization is tuned to maximize the phase modulation efficiency of the SLM). The first telescope (lenses L1 and L2) expands the beam-size to slightly overfill the active area of the SLM. The second telescope, in 4f configuration (lenses L3 and L4), projects the SLM onto the back-aperture of the objective while also resizing the beam to overfill the latter. The holographic nature of the trapping system enables movement of traps and manipulated objects not only in the transverse plane but also along the microscope's optical axis. A dichroic mirror (DM-IR, Fig. 1) allows simultaneous trapping and imaging by reflecting the trapping laser while transmitting the imaging excitation beams at shorter wavelengths. The same high NA objective is used for imaging as well as trapping.

*2.2 Materials*

LC samples are prepared between two glass substrates ~170 μm thick. The sample thickness is set by sandwiching silica microspheres with diameter 15 – 60 μm between these glass plates as they are sealed together by glue. The surface boundary conditions for the LC molecules on the glass substrates are set as follows: (a) treating the substrate with 2 wt.% aqueous solutions of N,N-dimethyl-n-octadecyl-3-aminopropyl-trimethoxysilyl chloride (Sigma Aldrich) sets **n**(**r**) perpendicular to the surface; (b) coating the substrate with polyimide (e.g., PI-2555, HD Microsystems) and rubbing the coated layer anchors the **n**(**r**) at the surface along the rubbing direction. A room-temperature nematic LC, 5CB (4-*pentyl*-4'-cyanobiphenyl, Frinton Labs, birefringence $\Delta n$=0.2) or ZLI-2806 (EM Chemicals, $\Delta n$=0.04) is doped with a chiral agent cholesteryl pelargonate (Sigma-Aldrich). The resultant pitch (*p*) of the cholesteric LC is determined by controlling the concentration of the chiral dopant in it (*p* is inversely proportional to the dopant's concentration). The cholesteric LCs chosen for the experiments described in this article have a pitch in the range from 0.5 μm to 20 μm. A dispersion of microparticles is prepared in the cholesteric LC by adding melamine resin spherical particles of 4 μm in diameter in the form of powder (obtained by drying aqueous dispersion of these particles from Fluka on a cover slide) and by sonicating the mixture to break apart colloidal aggregates. We also use Gallium Nitride (GaN) nanorods doped into the cholesteric LC as anisometrically shaped probes for LC manipulation and imaging [16]. The nanorods are ~10 μm in length and have a hexagonal cross-section with ~150 nm edge sides. The nanorods are grown on thin AlN buffer layers on Si(111) substrates by molecular beam epitaxy [41,42]. The nanorods are doped with Si such that the free-carrier concentration is around $1\times10^{18}$ cm$^{-3}$. At these doping densities, the optical absorption at 1064 nm is sufficiently low (~80 cm$^{-1}$) that heating of the LC while the rods are manipulated with focused laser beams is negligible. The long axis of the nanorods is along the c-axis (0001) of their wurtzite crystal structure, with the sidewalls being m-planes [(1 -1 0 0) family], producing the hexagonal cross-section. The nanorods are first dispersed in isopropanol and then transferred into the LC by mixing and letting isopropanol evaporate by heating the mixture to ~ 60 °C.

## 3. Results

### 3.1 Optical manipulation and nonlinear imaging of colloids in LC host media

HOT has found numerous applications in materials and biological sciences because of its ability to manipulate multiple objects simultaneously in 3D [43,44] as well as in allowing the generation of laser beams with optical vortices [45]. We demonstrate 3D manipulation and nonlinear imaging capabilities of the integrated setup by using an example of a system consisting of cholesteric LC as the host medium and colloids of spherical and anisometric (rod-like) shapes dispersed in it.

We illustrate manipulation of melamine resin (refractive index $n_p$=1.68) spherical microparticles dispersed in cholesteric LCs based on nematic 5CB (extra-ordinary refractive index, $n_e$=1.74, ordinary refractive index $n_o$=1.54) in a series of vertical cross-sectional images obtained in CARS and 3PEF (Fig. 2). The cell substrates are treated to provide strong planar surface boundary conditions. The particles have tangential anchoring on their surface because of which they cause distortion in the cholesteric layers. The particle-induced layer distortions are partially suppressed by the strong parallel boundary conditions at the cell substrates and depend on the particle size relative to the cholesteric pitch and cell thickness.

In general, motion of particles perpendicular to the layers is more difficult compared to that within a layer. The HOT system allows motion of particles along the optical axis of the microscope (direction $z$, Fig. 2a), i.e., across the cholesteric layers at relatively high values of optical power (~50 mW, compared to ~10 mW required to move the particles laterally). Due to the high NA of the objective used, the range of movement along $z$ attainable by HOT is limited to only a few microns. In the case of the images shown, a large range of particle displacements (tens of micrometers) along $z$ is achieved by use of HOT in conjunction with movement of the objective along $z$. The resultant configurations of cholesteric layers are imaged in 3PEF, Fig. 2(a-c), as well as in CARS, Fig. 2(d-f), in a different set of similar manipulations. The 3PEF images are obtained with excitation at 870 nm and detecting the fluorescence by use of a BPF 417/60 nm [46]. The 3PEF spectrum obtained from 5CB along with the excitation used is shown in Fig. 2g. The CARS signal is generated by mixing of laser pulses at different frequencies: the pump and the probe pulse at frequency $\omega_p$ and the Stokes pulse at $\omega_{Stokes}$ to give the output at $\omega_{CARS} = \omega_p + (\omega_p - \omega_{Stokes})$. The output is detected for $\omega_p - \omega_{Stokes} = \omega_{CN}$ (corresponding to the CN stretching vibration at 2236 cm$^{-1}$, which is along the long molecular axis of 5CB). For the set of wavelengths used for pump/probe (780 nm) and Stokes (broadband), the CARS images due to CN stretch-vibration are obtained for the signal centered at approximately 664 nm, which is selected by a BPF 661/20 nm. The broadband Stokes pulse excites multiple vibrational modes in the molecules, all of which show up in the CARS spectrum, as seen in Fig. 2h (molecular structure of 5CB is shown in the inset). Thus we obtain the full CARS spectral information about the material and can select a particular vibration of interest for imaging.

The shape-anisotropy of microparticles dispersed in LCs lends to them interesting trapping behavior, elasticity-mediated interactions and self-assembling properties [13,16]. We use such a soft-matter system of GaN nanorods ($n_{GaN} \approx 2.3$) dispersed in cholesteric LC to illustrate the capabilities of the integrated HOT and MNOPM setup. The GaN nanorods tend to align parallel to the local **n(r)**, as they set tangential alignment for the LC molecules. In a cholesteric LC, they are therefore parallel to the cholesteric layers. This allows them to be used effectively as probes of **n(r)** [16]. GaN has a wurtzite non-centrosymmetric crystal structure and hence generates a strong second-harmonic response dependent on the relative orientation of the nanorod to the polarization of the incident radiation – thus enabling their facile polarization-dependent nonlinear imaging. Figure 3 shows this combined nonlinear imaging, e.g., in Fig.3(a,b), the LC is imaged in 3PEF and a pair of nanorods in SHG mode. Figure 3(c) shows the overlapping spectra of the 3PEF signal from LC and the SHG peak from the nanorods (at 435 nm) for excitation beam at 870 nm; both are simultaneously collected by use of the same BPF 417/60 nm. Figure 3(d,e) shows similar set of images obtained with the LC imaged in CARS and the nanorods in SHG mode. It is clear from the cross-sections, Fig. 3(b,e), that the nanorod aligns parallel to the layers. The apparently increased thickness of the nanorod in the cross-sectional images is due to its thermal diffusion over the signal-integration period and also because of the diffraction-limited spatial resolution.

The strength of interactions between colloidal inclusions and LC line defects depends on whether the defect core is singular or not [16]. Here we show a nanorod weakly interacting with a dislocation with Burgers vector $b=p$, having its core split into two nonsingular disclinations, $\lambda^{-1/2}\lambda^{+1/2}$. Figure 3(f) shows a nanorod aligned parallel to the dislocation with core shown in a vertical cross-section in Fig. 3(g). The nanorod is also aligned along the core of the defect line shown in the cross-section in Fig. 3(h). The schematic representation of the director structure forming the dislocation and the position of the nanorod embedded in it is shown in Fig. 3(i). The nanorod (or an aggregate of multiple nanorods) can be manipulated via non-contact rotation and translation while the LC structure and the colloidal inclusion are simultaneously imaged in 3D. We have tested that the interaction of the two-rod aggregate with the $b=p$ dislocation is weak (elasticity-mediate forces <5 pN), which is natural, as it has a non-singular structure of the defect core [16]. Since $n(r)$ in the core of the $\lambda$-disclination is along the disclination, the nanorod prefers to align along the defect line.

*3.2 Optical generation of LC defect structures*

It has been long known that LCs respond to an applied electric field (the so-called Freederickz transition) [1]. This effect can also be observed in the case of an electric field of a laser beam focused into the LC sample. However, upon removal of the electric/optical field, the LC realigns to its previous equilibrium ground state. In the case of confinement-frustrated LCs though, stable equilibrium deformation states of $n(r)$ can be induced by application of external fields or by laser beams [20,27,30]. We show here an example of an optically generated and imaged loop of the so-called cholesteric finger of the first kind (CF-1) [30]. The equilibrium pitch of the cholesteric LC used is ~ 15 μm. The LC is filled in a wedge-shaped cell and the structures are generated at a place where the cell thickness is ~20 μm. The cell substrates are treated so as to set $n(r)$ at the surfaces perpendicular to the substrates (i.e., giving a strong homeotropic anchoring). The surface anchoring forces the LC molecules, against their inherent tendency to twist, to be aligned uniformly perpendicular to the cell substrates. This confinement-induced frustration can be locally relieved in the cell via generation of localized defect structures with a laser beam. The value of the pitch (relative to the thickness of the cell) and the type of the beam used (e.g., optical vortices that can be generated by the SLM based HOT setup), determine the nature (size and the topological skeleton) of the defect structure formed [27].

An optically generated loop of CF-1 is imaged in 3PEF mode of MNOPM, and the corresponding in-plane and vertical cross-sections are shown in Fig. 4(a,d). An identical structure is generated in similar fashion and imaged in CARS mode, Fig. 4(c,e). The basic structure of a CF-1 type finger consists of four non-singular disclination lines, two $\lambda^{+1/2}$ (red) and two $\lambda^{-1/2}$ (blue) defect lines, as shown in Fig. 4(b). The total topological charge due to the loops is thus conserved. The reconstruction of the axially symmetric $n(r)$ and the defect loops in the structure are shown as a schematic, Fig. 4(b).

## 4. Discussion

*4.1 Comparison of the integrated system with other imaging and trapping approaches*

Optical manipulation has proved to be a tool of immense utility in quantifying elasticity-mediated colloidal interactions and tension of defect lines in LCs [17-21]. Optical tweezers based on acousto-optic deflectors and scanning galvanomirrors allow for measurement of these forces, which are typically probed only within the microscope's lateral plane. HOT extends these capabilities of optical tweezers to facilitate manipulation along the microscope's axial direction, also allowing simultaneous multi-particle trapping [31]. Combining HOT with high-resolution 3D multi-particle tracking system enables precise real-time 3D localization of the trapped colloids [47]. This helps determine the map of interaction forces exerted on each particle in 3D. In addition to 3D manipulation capabilities, HOT allows for generation of beams with phase singularities, which play an important role in controlling the LC defects [27]. Optical trapping has been previously combined with nonlinear imaging by performing both the tasks using the same pulsed-laser – the laser beam that traps the object can be used to simultaneously image it in a nonlinear modality such as two-photon fluorescence [48], Raman micro-spectroscopy [49], or broadband supercontinuum CARS [50]. Although these integrated trapping and imaging techniques have been immensely useful in biophysics, they do not allow orientational imaging of long-range molecular patterns that are critically important when performing such manipulation in anisotropic LC fluids. These integrated imaging-trapping systems also lack multimodal nonlinear optical imaging capabilities, thus limiting the scope of their application in the study of LCs.

Even though FCPM has been successfully integrated with optical manipulation and used to image LC director patterns [13,17,27], it requires doping LCs with specific dyes. For some LC imaging applications, it is hard to find a suitable dye that would align both excitation and fluorescence transition dipoles along **n(r)**. Moreover, to image biaxial LCs with FCPM, one would be required to use different dye molecules aligning along the different biaxial directors [51], while MNOPM can perform such imaging by use of appropriate chemical bonds of the constituent molecules. Another direct advantage of using nonlinear imaging is the enhanced orientational sensitivity. The intensity of signal in all MNOPM imaging modalities depends on the relative angular separation ($\theta$) between the light polarization and the local **n(r)** at the point of focus (see Table 1). In the case of FCPM, this dependence is $\sim cos^2\theta$ for unpolarized detection and $\sim cos^4\theta$ for linearly polarized detection collinear with the polarization direction of the excitation light. The angular dependence of intensity is much stronger in all MNOPM modalities. In the case of the SHG and two-photon excitation fluorescence, the intensity in images scales as $\sim cos^4\theta$ for unpolarized and as $\sim cos^6\theta$ for collinear polarized detection. In the case of CARS and 3PEF that utilize the third-order nonlinear processes, the intensity scales as $\sim cos^6\theta$ (without a polarizer in the detection channel) and $\sim cos^8\theta$ (with the detection polarizer collinear with polarization of excitation beams). These stronger dependencies on $\theta$ compared to FCPM impart an enhanced sensitivity to spatial variations in **n(r)** in all MNOPM imaging modalities. FCPM imaging of LCs with twisted configurations of **n(r)**, with the helical axis parallel to the microscope's axis is difficult, because the polarization of the imaging light follows the twist of **n(r)** as it propagates through the LC. This so-called Mauguin effect is quantified by the Mauguin parameter, $\eta_{Mauguin} = p\Delta n/2\lambda$, where $p$ is the pitch of the twist in **n(r)** structure being imaged, $\Delta n$ is its birefringence and $\lambda$ is the wavelength of imaging light. Large values of $\eta_{Mauguin} \gg 1$ are detrimental to the FCPM imaging and preclude the reconstruction of **n(r)**. MNOPM partially mitigates this issue, as $\lambda$ is in the near-infrared range. The use of longer excitation wavelengths in MNOPM imaging also helps improve the penetration depth inside the sample compared to FCPM (Table 1). Furthermore, by using short-pitch cholesteric LCs and comparing the FCPM and MNOPM images of sub-micrometer cholesteric layered structures having both vertical and horizontal layers, we find that MNOPM imaging modalities offer somewhat better radial and axial spatial resolution, owing to inherently localized excitation of nonlinear optical signals (Table 1).

The strength of the MNOPM approach is fully realized when employed to study composite soft-matter systems with some of the constituents exhibiting long-range orientational order, such as LC colloidal dispersions. The PCF based implementation of broadband CARS [52,53] and other imaging modalities makes our system cost-effective compared to other implementations [54,55] because only one laser is used to generate all the needed excitation pulses. Furthermore, CARS imaging with a broadband Stokes pulse enables imaging over a broad spectral range (~ 800 $cm^{-1}$ - 3500 $cm^{-1}$), probing multiple vibration states at the same time. The integration of HOT with MNOPM greatly enhances the capabilities in optical control and characterization of soft-matter systems in an unprecedented manner, as both the components of the integrated system work simultaneously yet independently of each other.

*4.2 Mitigating artifacts in nonlinear optical imaging and trapping*

There are several artifacts one needs to be cautious about while imaging LCs. As discussed earlier, nonlinear imaging techniques mitigate the Mauguin effect because the excitation wavelengths are longer, but the imaging quality can still be adversely affected when LCs with high birefringence and very slowly twisting director are imaged. Another artifact that requires attention is unintended light induced realignment of **n(r)** by the imaging beam – the optical Freederickz transition. Optical realignment occurs above a certain threshold laser power determined by the cell thickness and the birefringence of the material. This limits the maximum intensity that can be used to image LCs and hence ultimately sets a limit on the speed of imaging [33]. The optical forces exerted by the imaging beam (both gradient and scattering forces) on the colloidal particles or director structures can be a significant source of unintended optical manipulation when high optical power and slow rate of beam scanning are used. For the imaging reported here, the optical power at the sample is small enough (combined power < 2 mW) that effects due to the optical Freederickz transition as well as those due to optical forces are negligible.

Similar artifacts may also adversely affect optical manipulation in LCs. In addition, sample heating by the trapping beam may change the pitch of the cholesteric LC being studied, cause local convective flows, or induce phase transitions. Therefore it is critical to ensure that the LC and the particles being manipulated have negligible absorption at the trapping wavelength. Optical trapping at powers above the realignment threshold is modified by the Freederickz transition. This adds elastic forces due to the induced **n(r)** deformation in conjunction with the optical gradient and scattering forces between manipulated objects and the laser trap [20]. Optical manipulation in LCs is further complicated by the fact that they are birefringent turbid media. Birefringence causes defocusing and depolarization as the beam is focused deep into the sample. Due to the optical anisotropy of LCs, trapping in LCs is polarization dependent [15] and therefore depolarization can affect force measurements. The light scattering due to thermal fluctuations of **n(r)** also reduces trapping efficiency. These effects due to birefringence, turbidity and the optical aberrations along the optical path can be accounted and compensated for in real-time with the same SLM used to generate the laser trap patterns [56].

## 5. Conclusion

We have demonstrated the first successful integration of multimodal nonlinear optical microscopy with 3D optical manipulation and its application to imaging of LCs. The combined capabilities of labeling-free, 3D optical imaging in CARS, 3PEF, SHG and other nonlinear optical modes, with non-contact 3D optical manipulation facilitate the design of novel composite soft-materials and chemical-sensitive characterization of the underlying self-organization processes. The application of the integrated system extends to the study of multitudes of other soft-matter systems involving nanoparticles, lipids, DNA and other biological materials-based LCs, polymer dispersed LCs, etc. These soft-matter systems form the basis for novel materials with intriguing mechanical properties [57] (artificial muscles, materials with electro-optically controlled elasticity, high-strength fibers, etc.) and photonic applications (metamaterials, LC-based tunable lasers, etc.). The integrated system also has immense potential in molecular and cellular biology [58-64], e.g., in determining the mechanism behind how lipid membranes organize into complex-shaped organelles [58,59], in tracking cellular processes such as metabolism [60], reconstructing cell lineage [62] by imaging different components of the cell in different nonlinear modalities, and transcending the diffraction limit in near-field scanning nonlinear optical microscopy [64].

## 6. Acknowledgements

This work was supported by the Renewable and Sustainable Energy Initiative, the University of Colorado Innovation Seed Grant Program, International Institute for Complex Adaptive Matter, and by NSF grants DMR-0820579, DMR-0844115, DMR-0645461, and DMR-0847782. We acknowledge discussions with Noel Clark, Paul Ackerman, David Engström, Zhiyuan Qi, Bohdan Senyuk, and Mike Varney.

## References

1. P.G. de Gennes, and J. Prost, The Physics of Liquid Crystals (Clarendon Press, Oxford 1993).
2. S. J. Woltman , D. G. Jay, and G. P. Crawford, "Liquid-crystal materials find a new order in biomedical applications," Nature Mater. **6**, 929-938 (2007).
3. M. Camacho-Lopez, H. Finkelmann, P. Palffy-Muhoray, and M. Shelley, "Fast liquid-crystal elastomer swims into the dark," Nature Mater. **3**, 307-310 (2004).
4. I. I. Smalyukh, J. Butler, J. D. Shrout, M. R. Parsek, and G. C. L. Wong, "Elasticity-mediated nematiclike bacterial organization in model extracellular DNA matrix," Phys. Rev. E **78**, 030701(R) (2008).
5. I.I. Smalyukh, O.V. Zribi, J.C. Butler, O.D. Lavrentovich, G.C.L. Wong, "Structure and Dynamics of Liquid Crystalline Pattern Formation in Drying Droplets of DNA," Phys. Rev. Lett. **96**, 177801 (2006).
6. J. M. Brake, M. K. Daschner, Y. Y. Luk, and N. L. Abbott, "Biomolecular Interactions at Phospholipid-Decorated Surfaces of Liquid Crystals," Science **302**, 2094-2097 (2003).
7. P. Poulin, H. Stark, T. C. Lubensky, and D. A. Weitz, "Novel Colloidal Interactions in Anisotropic Fluids," Science **275**, 1770-1773 (1997).
8. T. C. Lubensky, D. Petty, N. Currier, and H. Stark, " Topological defects and interactions in nematic emulsions," Phys. Rev. E **57**, 610 (1998).
9. D. G. Grier, "Optical tweezers in colloid and interface science," Curr. Opin. Colloid. Interface Sci. **2**, 264 (1997).
10. D.J. Stevenson, F. Gunn-Moore, and K. Dholakia, "Light forces the pace: optical manipulation for biophotonics," J. Biomed. Opt **15**, 041503 (2010).
11. M. Yada, J. Yamamoto, and H. Yokoyama, "Direct Observation of Anisotropic Interparticle Forces in Nematic Colloids with Optical Tweezers," Phys. Rev. Lett. **92**, 18550 (2004).
12. I. I. Smalyukh, et al., "Elasticity-Mediated Self-Organization and Colloidal Interactions of Solid Spheres with Tangential Anchoring in a Nematic Liquid Crystal," Phys. Rev. Lett. **95**, 157801 (2005).
13. C. Lapointe, T. Mason, and I. I. Smalyukh, "Shape-Controlled Colloidal Interactions in Nematic Liquid Crystals," Science **326**, 1083-1086 (2009).
14. M. Škarabot, et al., "Laser trapping of low refractive index colloids in a nematic liquid crystal," Phys. Rev. E **73**, 021705 (2006).
15. I.I. Smalyukh, A.V. Kachynski, A.N. Kuzmin, and P.N. Prasad, "Laser trapping in anisotropic fluids and polarization controlled particle dynamics," Proc. Nat. Acad. Sci. U.S.A. **103**, 18048-18053 (2006).
16. D. Engstrom, M. Persson, R. P. Trivedi, K. A. Bertness, M. Goksor, and I. I. Smalyukh, "Three-dimensional long-range order and defect core structures in anisotropic fluids probed by nanorods," submitted.
17. R. P. Trivedi, D. Engstrom, and I. I. Smalyukh, "Optical manipulation of colloids and defect structures in anisotropic liquid crystal fluids," Jour. Opt. (to be published, 2010)
18. Y. Ivashita, and H. Tanaka, "Optical Manipulation of Defects in a Lyotropic Lamellar Phase," Phys. Rev. Lett. **90**, 045501 (2003).
19. I. I. Smalyukh, A. N. Kuzmin, A. V. Kachynskii, P. N. Prasad, and O. D. Lavrentovich, "Optical trapping of colloidal particles and measurement of the defect line tension and colloidal forces in a thermotropic nematic liquid crystal," Appl. Phys. Lett. **86**, 021913 (2005).
20. I. I. Smalyukh, D. S. Kaputa, A. V. Kachynski, A. N. Kuzmin, and P. N. Prasad," Optical trapping of director structures and defects in liquid crystals using laser tweezers," Opt. Express **15**, 4359 (2007).
21. I. I. Smalyukh, et al., "Optical Trapping, Manipulation, and 3D Imaging of Disclinations in Liquid Crystals and Measurement of their Line Tension," Mol. Cryst. Liq. Cryst. 450 79 (2006).

22. S. Wurlitzer, C. Lautz, M. Liley, C. Duschl, and T. M. Fischer, "Micromanipulation of Langmuir-Monolayers with Optical Tweezers," J. Phys. Chem. B **105**, 182 (2001).
23. N. Murazawa, S. Juodkazis, and H. Misawa, "Laser manipulation based on a light-induced molecular reordering," Opt. Express **14**, 2481 (2006).
24. E. Brasselet, T. Balcinuas, N. Murazawa, S. Juodkazis, and H. Misawa, "Light-Induced Nonlinear Rotations of Nematic Liquid Crystal Droplets Trapped in Laser Tweezers," Mol. Cryst. Liq. Cryst. **512**, 143-151(2009).
25. H. F. Gleeson, T. A. Wood, and M. Dickinson, "Laser manipulation in liquid crystals: an approach to microfluidics and micromachines," Phil. Trans. R. Soc. A **364** 2789-2805 (2006).
26. H. Stark, "Physics of colloidal dispersions in nematic liquid crystals," Phys. Rep. **351** 387-474 (2001).
27. I. I. Smalyukh, Y. Lansac, N. A. Clark, and R. P. Trivedi, "Three-dimensional structure and multistable optical switching of triple-twisted particle-like excitations in anisotropic fluids," Nature Mater. **9** 139-145 (2010).
28. I.I. Smalyukh, S.V. Shiyanovskii, and O.D. Lavrentovich, "Three-dimensional imaging of orientational order by fluorescence confocal polarizing microscopy," Chem. Phys. Lett. **336**, 88-96 (2001).
29. I. I. Smalyukh, and O. D. Lavrentovich, "Three-dimensional director structures of defects in Grandjean-Cano wedges of cholesteric liquid crystals studied by fluorescence confocal polarizing microscopy," Phys. Rev. E **66**, 051703 (2002).
30. I.I. Smalyukh, et al., "Electric-field-induced nematic-cholesteric transition and three-dimensional director structures in homeotropic cells," Phys. Rev. E **72**, 061707 (2005).
31. S. Anand, R. P. Trivedi, G. Stockdale, and I. I. Smalyukh, "Non-contact optical control of multiple particles and defects using holographic optical trapping with phase-only liquid crystal spatial light modulator," Proc. SPIE **7232** 723208 (2009).
32. A.V. Kachynski, A.N. Kuzmin, P.N. Prasad, and I.I. Smalyukh, "Coherent anti-Stokes Raman scattering polarized microscopy of 3-D director structures in liquid crystals," Appl. Phys. Lett. **91**, 151905 (2007).
33. A. V. Kachynski, A. N. Kuzmin, P. N. Prasad, and I. I. Smalyukh, "Realignment-enhanced coherent anti-Stokes Raman scattering and three-dimensional imaging in anisotropic fluids," Optics Express **16**, 10617-10632 (2008).
34. B. G. Saar, H.-S. Park, X. S. Xie, and O. D. Lavrentovich, "Three-dimensional imaging of chemical bond orientation in liquid crystals by coherent anti-Stokes Raman scattering microscopy," Opt. Express **15**, 13585-13596 (2007).
35. B.-C. Chen, and S.-H. Lim, "Three-dimensional imaging of director field orientations in liquid crystals by polarized four-wave mixing microscopy," Appl. Phys. Lett. **94**, 171911 (2009).
36. K. Yoshiki, M. Hashimoto, and T. Araki, "Second-Harmonic-Generation Microscopy Using Excitation Beam with Controlled Polarization Pattern to Determine Three-Dimensional Molecular Orientation," Japan. J. Appl. Phys. **44** 1066-1068 (2005).
37. R.S. Pillai, M. Oh-e, H. Yokoyama, C.J. Brakenhoff, and M. Muller, "Imaging colloidal particle induced topological defects in a nematic liquid crystal using third harmonic generation microscopy," Opt. Express **14**, 12976-12983 (2006).
38. D. A. Higgins, and B. J. Luther, "Watching molecules reorient in liquid crystal droplets with multiphoton-excited fluorescence microscopy," J. Chem. Phys. **119**, 3935 (2003).
39. T. Lee, R. P. Trivedi, and I. I. Smalyukh, "Multimodal nonlinear optical polarizing microscopy of long-range molecular order in liquid crystals," Opt. Lett. **35**, 3447-3449 (2010).
40. The full description of the procedures used in this paper requires the identification of certain commercial products and their suppliers. The inclusion of such information should in no way be construed as indicating that such products or suppliers are endorsed by NIST or are recommended by NIST or that they are necessarily the best materials, instruments, or suppliers for the purposes described.
41. K. A. Bertness, et al., "Mechanism for spontaneous growth of GaN nanowires with molecular beam epitaxy", J. Cryst. Growth **310**, 3154-3158 (2008).
42. J. B. Schlager, K. A. Bertness, P. T. Blanchard, L. H. Robins, A. Roshko, and N. A. Sanford, "Steady-state and time-resolved photoluminescence from relaxed and strained GaN nanowires grown by catalyst-free molecular-beam epitaxy," J. Appl. Phys. **103**, 124309 (2008).
43. J. Liesener, M. Reicherter, T. Haist, and H. J. Tiziani, "Multi-functional optical tweezers using computer-generated holograms," Opt. Commun. **185**, 77–82 (2000).
44. J. E. Curtis, B. A. Koss, and D. G. Grier, "Dynamic holographic optical tweezers," Opt. Commun. **207**, 169–175 (2002).
45. J. E. Curtis, and D. G. Grier, "Structure of optical vortices," Phys. Rev. Lett. **90**, 133901 (2003).
46. The notation such as BPF 417/60 denotes a bandpass filter with center wavelength of 417nm and 60nm bandwidth.
47. D. B. Conkey, et al., "Three-dimensional parallel particle manipulation and tracking by integrating holographic optical tweezers and engineered point spread functions," Opt. Exp. (to be published, 2010).
48. K. Dholakia, et al., "Imaging in optical micromanipulation using two-photon excitation," New. J. Phys. **6**, 136 (2004).
49. P. R. T. Jess, at al., "Simultaneous Raman micro–spectroscopy of optically trapped and stacked cells," J. Raman Spectrosc. **38**, 1082 (2007).

50. K. Shi, P. Li, and Z. Liu, "Broadband coherent anti-Stokes Raman scattering spectroscopy in supercontinuum optical trap," Appl. Phys. Lett. **90**, 141116 (2007).
51. I.I. Smalyukh, R. Pratibha, N.V. Madhusudana, and O.D. Lavrentovich, "Selective imaging of 3-D director fields and study of defects in biaxial smectic A liquid crystals," Eur. Phys. J. E **16**, 179-192 (2005).
52. T. W. Kee, an M. T. Cicerone, "Simple approach to one-laser, broadband coherent anti-Stokes Raman scattering microscopy," Phys. Rev. Lett. 29, 2701 (2004).
53. H. Kano, and H. Hamaguchi, "Femtosecond coherent anti-Stokes Raman scattering spectroscopy using supercontinuum generated from a photonic crystal fiber," Appl. Phys. Lett. **85**, 4298 (2004).
54. J. X. Cheng, and X. S. Xie, " Coherent anti-Stokes Raman scattering microscopy: Instrumentation, Theory, and applications," J. Phys. Chem B 108, 827-840 (2004).
55. F. Ganikhanov, S. Carrasco, X. S. Xie, M. Katz, W. Seitz, and D. Kopf, "Broadly tunable dual-wavelength light source for coherent anti-Stokes Raman scattering microscopy," Opt. Lett. **31**, 1292-1294 (2006).
56. T. Cizmar, M. Mazilu, and K. Dholakia, "In situ wavefront correction and its application to micromanipulation," Nature Photon. 4, 388 (2010)
57. M. Zapotocky, L. Ramos, P. Poulin, T. C. Lubensky, and D. A. Weitz, "Particle-Stabilized Defect Gel in Cholesteric Liquid Crystals," Science 283, 209 (1999).
58. G. K. Voeltz, and W. A. Prinz, "Sheets, ribbons and tubules — how organelles get their shape," Nature Rev. Molec. Cell Biol. 8, 258-264 (2007).
59. Y. Shibata Y, G. K. Voeltz, and T. A. Rapoport, "Rough sheets and smooth tubules," Cell **126**, 435-9 (2006).
60. X. S. Xie, J. Yu, and W. Y. Yang, "Living Cells as Test Tubes", Science **312**, 228 (2006).
61. Y. Fu, H. Wang, R. Shi, and J.-X. Cheng, "Second Harmonic and Sum Frequency Generation Imaging of Fibrous Astroglial Filaments in Ex Vivo Spinal Tissues," Biophysical Journal 92, 3251 (2007)
62. N. Olivier, et al., "Cell Lineage Reconstruction of Early Zebrafish Embryos Using Label-Free Nonlinear Microscopy," Science **329**, 967 (2010).
63. H. Chen, et al., "A multimodal platform for nonlinear optical microscopy and microspectroscopy," Opt. Express **17**, 1282 (2009).
64. R. D. Schaller, et al., "Nonlinear Chemical Imaging Nanomicroscopy: From Second and Third Harmonic Generation to Multiplex (Broad-Bandwidth) Sum Frequency Generation Near-Field Scanning Optical Microscopy," Jour. Phys. Chem. B **106**, 5143 (2002).

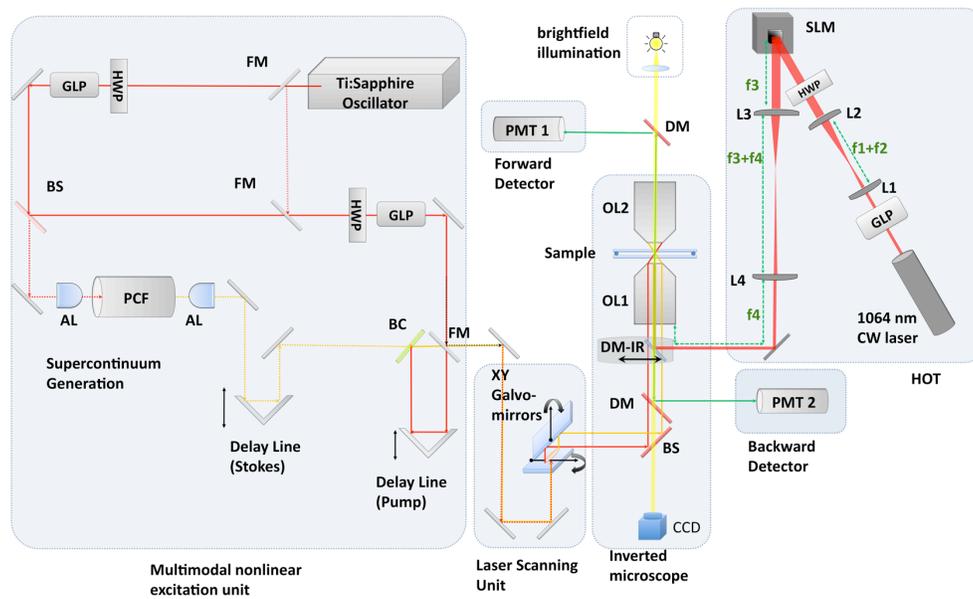

Figure 1. Schematic diagram of the MNOPM and HOT optical setup. The integrated MNOPM and HOT setup is built around an inverted microscope. The components of MNOPM: HWP: half wave-plate, GLP: Glan-laser polarizer, PCF: highly nonlinear photonic crystal fiber, AL: aspheric lens, BS: beam-splitter, BC: beam combiner, FM: folding mirror, PMT: photo-multiplier tube, DM: dichroic mirror, OL: objective lens. The components of the HOT: SLM: spatial light modulator, L1, L2, L3, L4: plano-convex lenses, DM-IR: dichroic mirror for the IR trapping laser.

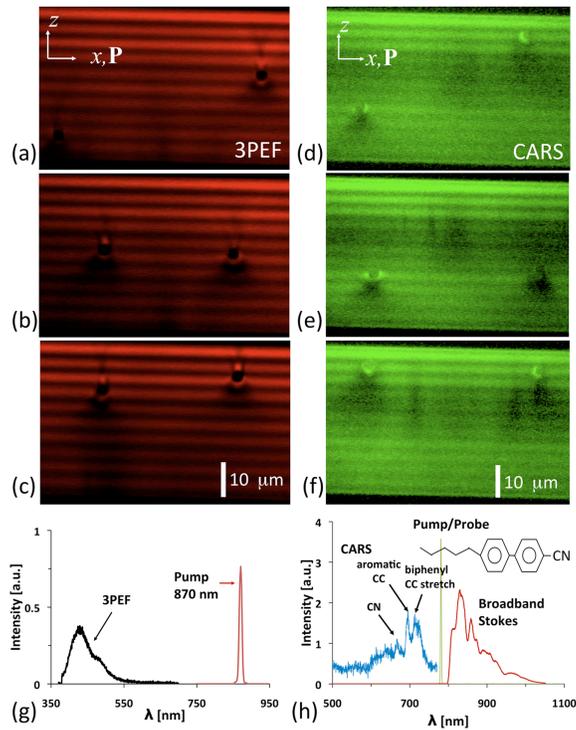

Figure 2. Manipulation and imaging of spherical microparticles in a cholesteric LC. Vertical cross sectional images of planar cholesteric LC cells obtained in 3PEF (a-c) and CARS (d-f) modes and showing the lamellar structure with spherical microparticles embedded in it. Particles of 4 μm diameter are optically trapped and moved within as well as across the cholesteric layers. The nonlinear signals are collected at the forward detection with a BPF 417/60 for 3PEF and a BPF 661/20 for CARS of CN stretching vibration of 5CB. (g) shows the excitation and the fluorescence spectrum of 3PEF for 5CB. (h) shows the broadband Stokes, the pump at 780 nm and the broadband CARS spectrum obtained from the LC (with different peaks in the CARS spectrum marked to indicate the corresponding Raman vibration). Molecular structure of 5CB is shown in the inset.

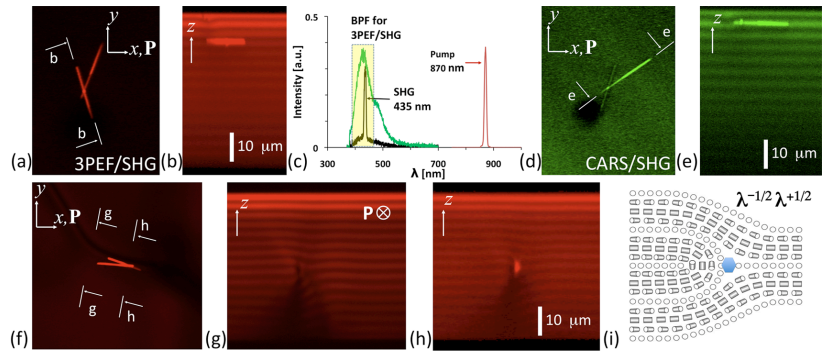

Figure 3. 3D Imaging of GaN nanorods dispersed in a cholesteric LC. GaN nanorods are manipulated in the layered structure of the cholesteric LC and the LC-nanorod composite is imaged in its lateral plane as well as in the vertical cross-section in two nonlinear modes: the LC is probed using 3PEF (a, b) and CARS (d, e) and the nanorods are visualized in SHG. (c) shows the spectrum with the excitation at 870 nm and the 3PEF emission from the LC molecules and SHG (at 435 nm) from the GaN nanorods. The nanorods interacting with a cholesteric dislocation are imaged in 3PEF (f-h). (g) shows 3PEF image of a dislocation comprising $\lambda^{+1/2}$ and $\lambda^{-1/2}$ disclinations in its vertical cross-section. The schematic diagram of the molecular director pattern around the dislocation with the position of the nanorod embedded in it is shown in (i).

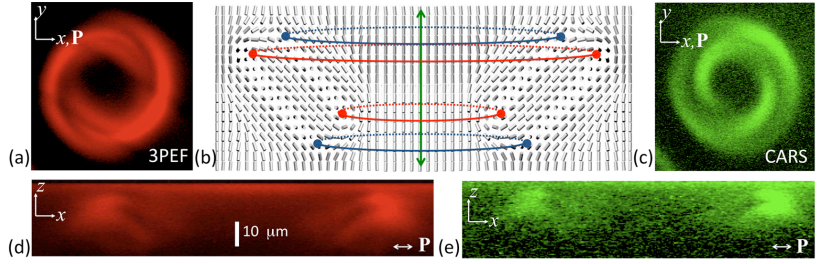

Figure 4. 3D imaging of optically generated defect structures in a homeotropic unwound cholesteric LC sample. Loops of cholesteric finger CF1 are optically generated with the HOT setup and are imaged in their lateral and vertical cross-sectional planes, in 3PEF (a, d) and CARS (c, e) imaging modes. The schematic representation of the twist of $\mathbf{n}(\mathbf{r})$ within the structure is shown in its vertical cross section (b). The red dots and lines represent the loop of nonsingular disclinations $\lambda^{+1/2}$ while the blue dots and lines correspond to $\lambda^{-1/2}$ disclination loops. The structure is axially symmetric around the axis marked by the green double arrow.

Table 1. Comparison of different orientation-sensitive 3D microscopy techniques

| Microscopy | 3D spatial resolution | | Dye labeling | Orientational sensitivity | | Mauguin effect | Penetration depth | |
|---|---|---|---|---|---|---|---|---|
| | radial | axial | | unpolarized detection | polarized detection | | 5CB | ZLI-2806 |
| FCPM | 0.25μm | 0.55μm* | Yes | $\cos^2\theta$ | $\cos^4\theta$ | stronger | ~20μm | ~80μm |
| SHG & 2PEF (of MNOPM) | 0.2μm | 0.4μm | Yes/No | $\cos^4\theta$ | $\cos^6\theta$ | weaker | ~100μm | ~200μm |
| 3PEF (of MNOPM) | 0.2μm | 0.4μm | Yes/No | $\cos^6\theta$ | $\cos^8\theta$ | weaker | ~150μm | ~200μm |
| CARS (of MNOPM) | 0.2μm | 0.4μm | No | $\cos^6\theta$ | $\cos^8\theta$ | weaker | ~150μm | ~200μm |

* Measured in low-birefringence LC, ZLI-2806 (much worse for high-birefringence LCs)